\documentclass[prl,aps,
twocolumn,
nofootinbib,
floatfix,
superscriptaddress]{revtex4}
\usepackage{graphicx}
\usepackage{amsmath,amssymb,graphicx}
\usepackage{multirow}
\usepackage{slashbox}
\usepackage{longtable}
\usepackage{color}

\def\aa1{\phi}
\def\cc1{\psi}

\catcode`\@=12

\begin{document}


\title{\bf On the Regularization of On-Shell Diagrams}


\author{
\scshape Paolo Benincasa${}^{\dagger}$, Eduardo Conde${}^{\ddagger}$,
         David Gordo${}^{\dagger}$\\[0.4cm]
\ttfamily ${}^{\dagger}$Instituto de F{\'i}sica Te{\'o}rica, \\
\ttfamily Univerisdad Aut{\'o}noma de Madrid / CSIC \\
\ttfamily Calle Nicolas Cabrera 13, Cantoblanco 28049, Madrid, Spain\\
\small \ttfamily paolo.benincasa@csic.es, david.gordo@uam.es \\[0.2cm]
\ttfamily ${}^{\ddagger}$Service de Physique Th{\'e}orique et Math{\'e}matique,\\
\ttfamily Universit{\'e} Libre de Bruxelles and International Solvay Institutes,\\
\ttfamily Campus de la Plaine, CP 231, B-1050 Bruxelles, Belgium\\
\small \ttfamily econdepe@ulb.ac.be
}

\begin{abstract}
 In this letter we discuss a regularization scheme for the integration of 
generic on-shell forms. The basic idea is to extend the three-particle 
amplitudes to the space of unphysical helicities keeping the dimension of 
the related coupling constant fixed, and construct on-shell forms out of 
them. We briefly discuss the analytic structure of the extended on-shell
diagrams, both at tree level and one loop.
Furthermore, we propose an integration contour which, applied to the
relevant on-shell forms, allows to extract the four-particle amplitudes in 
Lorentz signature at one loop. With this contour at hand, we explicitly apply our procedure
to this case obtaining the IR divergencies as poles in the deformation 
parameter space, as well as the correct functional form for the finite
term. This procedure provides a natural regularization for 
generic on-shell diagrams.
\end{abstract}

\maketitle

\section{Introduction}\label{sec:Intro}

In recent years the development of on-shell methods to compute scattering 
amplitudes has led to a new way of thinking of perturbation theory.
In the regime in which asymptotic states can be defined, it is possible to 
formulate a theory
in terms of on-shell processes from first principles, with no reference to
ia pre-existent Lagrangian. This program has been extensively developed for
planar $\mathcal{N}\,=\,4$ Super Yang-Mills theory in four dimensions 
and $\mathcal{N}\,=\,6$ ABJM \cite{ArkaniHamed:2012nw}, leading to 
the formulation of scattering processes first in terms of a(n auxiliary) 
Grassmannian \cite{ArkaniHamed:2012nw, Huang:2013owa, 
Huang:2014xza, Elvang:2014fja} as well as, just in the case of planar
$\mathcal{N}\,=\,4$ SYM, of a geometrical object, the {\it amplituhedron}, 
from which the amplitudes can be read off as volumes
\cite{Arkani-Hamed:2013jha}.

In a nutshell, the general idea underlying these constructions 
is to build-up on-shell processes
by suitably gluing the three-particle amplitudes: one imposes momentum
conservation on the internal states 
keeping them on-shell, and integrating over their phase-space 
\cite{ArkaniHamed:2012nw}. 
Depending on how many legs are glued together and 
how many three-particle amplitudes are involved, the gluing procedure can 
fix all the degrees of freedom on the internal legs,
or some of them may stay unfixed, 
or also constraints can be imposed on the external momenta.

In this language, the perturbative expansion of a scattering amplitude 
in four-dimensions is represented as a series of $4L$-forms
\begin{equation}\label{eq:AmplExp}
 \mathcal{M}_n(\{p_{i},\,h_i\})\,=\,
  \sum_{L\,=\,0}^{\infty}\mathcal{M}_{n}^{\mbox{\tiny $(L)$}}(\{p_i,\,h_i\},\,\{z_l\})\bigwedge_{l=1}^{4L}dz_{l},
\end{equation}
where $h_i$'s are the helicities of the external states, 
$z_l$ ($l\,=\,1,\ldots,\,L$) are the unfixed degrees of freedom left over by 
the gluing procedure which generates
the on-shell diagrams at a given order $L$, and the coefficients 
$\mathcal{M}_{n}^{\mbox{\tiny $(L)$}}(\{p_i,\,h_i\},\,\{z_l\})$ are rational functions of both the momenta of the external states $p_i$ and of $z_l$. 
Notice that $\mathcal{M}_{n}^{\mbox{\tiny $(L)$}}$ can, in principle, 
contain terms with different powers of the coupling constant -- for example,
the $L=0$ term contains both the tree-level amplitude and further rational 
terms coming from higher loops.
At a given order $L$, the parameters $z_l$ form a $4L$-dimensional phase
space which turns out to be equivalent to the loop momenta phase-space at $L$-loops in the standard Feynman diagram
representation. Thus, the on-shell diagrammatics involves the {\it integrands} and the rational terms characterizing
an amplitude rather than the {\it integrated} amplitude. In order to obtain 
an actual physical amplitude at a given 
order $L$, the relevant on-shell diagrams
need to be integrated over a suitable contour, but it is currently not known
how to identify the one which provides the Lorentz-signature amplitude.
This issue 
comes necessarily with the question of how to regularize the IR-divergences
which plague massless theories.



A first approach to this issue for $\mathcal{N}\,=\,4$ SYM amplitudes was 
taken in \cite{Bourjaily:2013mma},
where the regulator was introduced by making the external states slightly
off-shell, and in
\cite{Ferro:2012xw, Ferro:2013dga}, where using integrability arguments
the on-shell diagrams were deformed by introducing spectral parameters which
are supposed to play the role of regulators. In the latter case, 
the integration of
the relevant on-shell diagram was performed by going back to the
off-shell loop momentum and choosing the parameters in such a way
that the resulting integrand was simply the original loop integrand with
the powers of the propagators analytically continued to complex values.
As proposed, this approach 
breaks Yangian symmetry, while demanding Yangian symmetry to be preserved 
beyond MHV level makes the deformation become trivial 
\cite{Beisert:2014qba}. A way out has
been argued to be the implementation of the spectral deformation directly
on the Grassmannian formulation of the theory \cite{Ferro:2014gca}: 
while at undeformed level the two representations are equivalent, the 
deformed Grassmannian does not have a direct on-shell diagrammatic 
interpretation. 

In this paper, we consider the on-shell diagrammatics as a 
more general framework and introduce a regularization prescription
which can be performed for a general theory, treating
directly the integration without any need of referring to any 
off-shell loop momenta.


\section{Continuation of the helicity space and locality}
\label{sec:ContHelSpace}
An S-matrix theory can be defined through the 
fundamental symmetries which one wants to attribute to it,
and its fundamental physical objects which are determined by such 
symmetries. For asymptotically Minkowski space-times in
the regime in which asymptotic states can be defined, Poincar{\'e} 
symmetry fixes
through the space-time translations and the Lorentz little group the 
three-particle amplitudes for massless theories
\cite{Benincasa:2007xk}
\begin{equation}\label{eq:3ptAmpl}
 M_3\:=\:
  \left\{
   \begin{array}{l}
    \kappa\,\langle1,2\rangle^{d_3}\langle2,3\rangle^{d_1}\langle3,1\rangle^{d_2},\hspace{.5cm}\sum_{i=1}^{3}h_i\,<\,0\\
    \phantom{\ldots}\\
    \kappa\,[1,2]^{-d_3}[2,3]^{-d_1}[3,1]^{-d_2},\:\sum_{i=1}^3h_i\,>\,0
   \end{array}
  \right.
\end{equation}
with the momenta represented as direct products of $SL(2)$ spinors
$p_{a\dot{a}}\,=\,\lambda_{a}\tilde{\lambda}_{\dot{a}}$, 
$[\cdot,\cdot]$ ($\langle\cdot,\cdot\rangle$) being the Lorentz invariant
internal product for (un)dotted spinors, 
$d_i\,=\,h_i-h_{i+1}-h_{i-1}$ and the dimension of the coupling 
constant $\kappa$ being $[\kappa]\,=\,1-|h_1+h_2+h_3|$. The latter can be 
used to classify theories \cite{Benincasa:2011pg}: a class
is identified by fixing the dimension of the coupling constant, which 
constrains the possible helicity configurations allowed. 
The expression (\ref{eq:3ptAmpl}) is non-trivial either if we 
consider the complexified Lorentz group $SO(3,1;\mathbb{C})$ and, thus,
complex momenta, or if the Lorentz group is taken to be $SO(2,2)$.
Diagrammatically, the three-particle amplitudes are represented by
a black (white) vertex for $h_1+h_2+h_3\,<\,0$ ($>\,0$) with incoming
(outgoing) arrows for negative (positive) helicity states:
\begin{figure}[htbp]
 \centering 
  \scalebox{.60}{\includegraphics{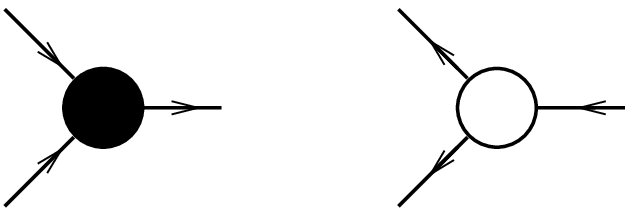}}
\end{figure}

Let us now consider the three-particle amplitudes (\ref{eq:3ptAmpl}), fix the dimension of the coupling constant, and
perform an extension of the helicity space to complex values
\begin{equation}\label{eq:helext}
 h_i\:\longrightarrow\:\mathfrak{h}_i\:\equiv\:h_i+\sigma_i\varepsilon_i,
 \qquad\sigma_{i}=-\frac{h_i}{|h_i|}
\end{equation}
in such a way that the dimension of the coupling constant is unchanged 
(notice that the choice of $\sigma_i$ is just a matter of convention and
it does not imply any assumption about the real part of $\varepsilon_i$). 
This requirement constrains the sum of the parameters
$\varepsilon_i$ (weighted by $\sigma_i$) to vanish 
and allows to extend a given 
class of four-dimensional theories to the unphysical helicity space. 
The extended three-particle amplitudes acquire the following form
\begin{equation}\label{eq:3ptAmplext}
 M_3(\varepsilon)\:=\:
  M_3\times\left\{
   \begin{array}{l}
    \frac{\langle2,3\rangle^{2\sigma_1\varepsilon_1}
          \langle3,1\rangle^{2\sigma_2\varepsilon_2}}{
        \langle1,2\rangle^{2(\sigma_1\varepsilon_1+\sigma_2\varepsilon_2)}},
     \quad\sum_{i=1}^{3}h_i\,<\,0\\
    \phantom{\ldots}\\
    \frac{[1,2]^{2(\sigma_1\varepsilon_1+\sigma_2\varepsilon_2)}}{
          [2,3]^{2\sigma_1\varepsilon_{1}}[3,1]^{2\sigma_2\varepsilon_2}},\;
     \quad\sum_{i=1}^3h_i\,>\,0
   \end{array}
  \right.
\end{equation}
where the constraint on the parameters $\varepsilon_i$ has been used to 
solve $\varepsilon_3$ as a function of the other two parameters.

With this extension at hand, we can build generalized on-shell diagrams by gluing the extended three-particle amplitudes
(\ref{eq:3ptAmplext}) 
as prescribed for the undeformed case. First of all, the
constraint on the parameters $\varepsilon_i$ of the 
three-particle amplitudes generalizes to the parameters associated
to the external states of any $n$-particle on-shell diagram
\begin{equation}\label{eq:nhelext}
 \sum_{i=1}^n\sigma_i\varepsilon_i\:=\:0.
\end{equation}
Secondly, the equivalence relation named {\it merger} 
\cite{ArkaniHamed:2012nw}, which connects two on-shell diagrams
made up by two three-particle amplitudes of the same type but glued
along different channels, still holds because it is just a consequence of
the proportionality of all the spinors of the same type.

Let us now consider the following extended on-shell diagram
\begin{equation}\label{eq:M4tree}
 \begin{split}
  {\raisebox{-0.75cm}{\scalebox{.40}{\includegraphics{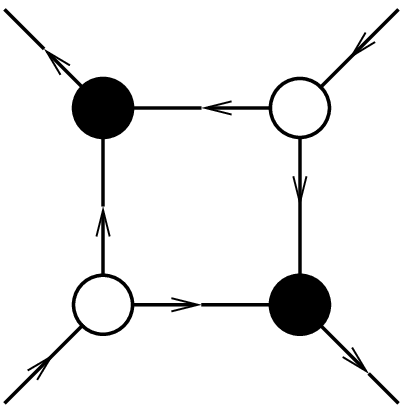}}}}
  \hspace{-.2cm}=\:
  M_4^{\mbox{\tiny tree}}
&\left(\frac{\langle2,3\rangle}{\langle1,3\rangle}\right)^{2\varepsilon_{12}}
\left(\frac{\langle1,2\rangle}{\langle1,3\rangle}\right)^{2\varepsilon_{23}}
 \times\\
 &\times
\left(\frac{\langle4,1\rangle}{\langle3,1\rangle}\right)^{2\varepsilon_{34}}
\left(\frac{\langle3,4\rangle}{\langle3,1\rangle}\right)^{2\varepsilon_{41}},
 \end{split}
\end{equation}
where $\varepsilon_{i,i+1}$ is the deformation parameter related to
the internal states between the particles labelled by $i$ and $i+1$,
$M_4^{\mbox{\tiny tree}}$ is, generically, a contribution to the tree-level
$4$-particle amplitude (or the full amplitude in presence of color ordering)
with the helicity configuration for the external states given on
the l.h.s of (\ref{eq:M4tree}).
Furthermore, the constraints on the deformation 
parameters have been used to express the external ones in terms of the
internal $\varepsilon_{i,i+1}$'s. One comment is now in order. Even
at tree level, our procedure breaks locality either enhancing already 
existent singularities or generating new ones. For configurations which
at $\varepsilon\,=\,0$ admit several representations, and thus further
equivalence relations, our deformation may {\it in general} break such
equivalences\footnote{In principle, one can make a non-trivial choice
of parameters in order to restore these equivalence relations at the price
of having less independent parameters. In this
case, locality still is broken but in a much milder way 
\cite{Benincasa:2014wp1}.}\cite{Benincasa:2014wp1}. 

To which extent is this
a big issue? For tree-level results this is not really relevant for
two reasons: a naive one is that they are reproduced setting the
parameters $\varepsilon_{i,i+1}$ exactly to zero. A more subtle one is 
related to the analytic structure of (\ref{eq:M4tree}): taking properly a 
collinear limit on it, the related singularity appears
as a simple pole $1/\varepsilon$ whose residue reproduces the 
correct factorization. A neat way to see this
is to consider a BCFW deformation \cite{Britto:2005aa} of (\ref{eq:M4tree}).
Taking for simplicity the case of pure Yang-Mills, let us perform the
BCFW shift $\tilde{\lambda}_1(w)=\tilde{\lambda}_1-w\tilde{\lambda}_2$, 
$\lambda_2(w)=\lambda_2+w\lambda_1$, and consider the following 
integration
\begin{equation}\label{eq:collim}
 \oint_{\gamma}\frac{dw}{w}M_4^{\mbox{\tiny tree}}(\varepsilon;w)\:=\:
 \mathcal{F}(\varepsilon)
 \frac{\hat{M}_3^{\mbox{\tiny H}}
  \hat{M}_3^{\mbox{\tiny A}}}{\langle2,3\rangle[1,3]}
 \oint_{\gamma}\frac{dw}{w}(w-w_{23})^{2\varepsilon_{12}-1},
\end{equation}
where $M_4^{\mbox{\tiny tree}}(\varepsilon;w)$ represents the rhs of 
(\ref{eq:M4tree}), $\mathcal{F}(\varepsilon)$ is all the 
$\varepsilon$-dependent factor of (\ref{eq:M4tree}) which stays 
$w$-independent, $w_{23}\,\equiv\,-\langle2,3\rangle/\langle1,3\rangle$,
the remaining $w$-independent factor comes just from the BCFW representation
of $M_{4}^{\mbox{\tiny tree}}$ (multiplied by $w_{23}$) of the r.h.s. of
(\ref{eq:M4tree}), and $\gamma$ is a closed path encircling $w_{23}$.
This is equivalent to analyzing one of the complex collinear limits in the
$t$-channel.
Expanding the integrand of (\ref{eq:collim}) in a neighborhood of $w_{23}$, 
taking its primitive and taking the limit 
$\varepsilon_{i,i+1}\,\longrightarrow\,0$ ($\forall\,i$)
in such a way that all the parameters go to zero in the same way, one gets
\begin{equation}\label{eq:collim2}
 \begin{split}
  &\lim_{\langle2,3\rangle\longrightarrow 0}
  P^2_{23}
  \oint_{\gamma}\frac{dw}{w}M_4^{\mbox{\tiny tree}}(\varepsilon;w)
   \:=\\
  &\hspace{.6cm}=\:
   \frac{1}{2\varepsilon_{12}}M_{3}^{\mbox{\tiny H}}(4,1,P_{23})
   M_{3}^{\mbox{\tiny A}}(-P_{23},2,3)+\mathcal{O}(\varepsilon^0).
 \end{split}
\end{equation}
In this specific case, it is actually the full tree-level amplitude which
appears as the residue of the pole in the $\varepsilon$-parameter space 
(before taking the limit $\langle2,3\rangle\,\longrightarrow\,0$).

At loop level the physical singularities get enhanced but also new 
({\it unphysical}) ones can appear. The first effect is just an analytic 
continuation of the powers of the loop propagators, while the 
new singularities just provide scheme-dependent quantities. For a deeper and
more extensive discussion about the helicity continuation and the effect
of locality breaking, we refer to \cite{Benincasa:2014wp1}.


\section{Helicity continuation and IR regularization}\label{sec:IRreg}

Let now explore the possibility that the parameters $\varepsilon_{i,i+1}$
can actually work as IR regulators. For this purpose, let us consider a 
contribution to a one-loop amplitude which contains just IR singularities
({\it i.e.} we consider the on-shell representative of a contribution
to a scalar box integral):
\begin{widetext}\begin{equation}\label{eq:M41L}
 \begin{split}
  {\raisebox{-1cm}{\scalebox{.25}{\includegraphics{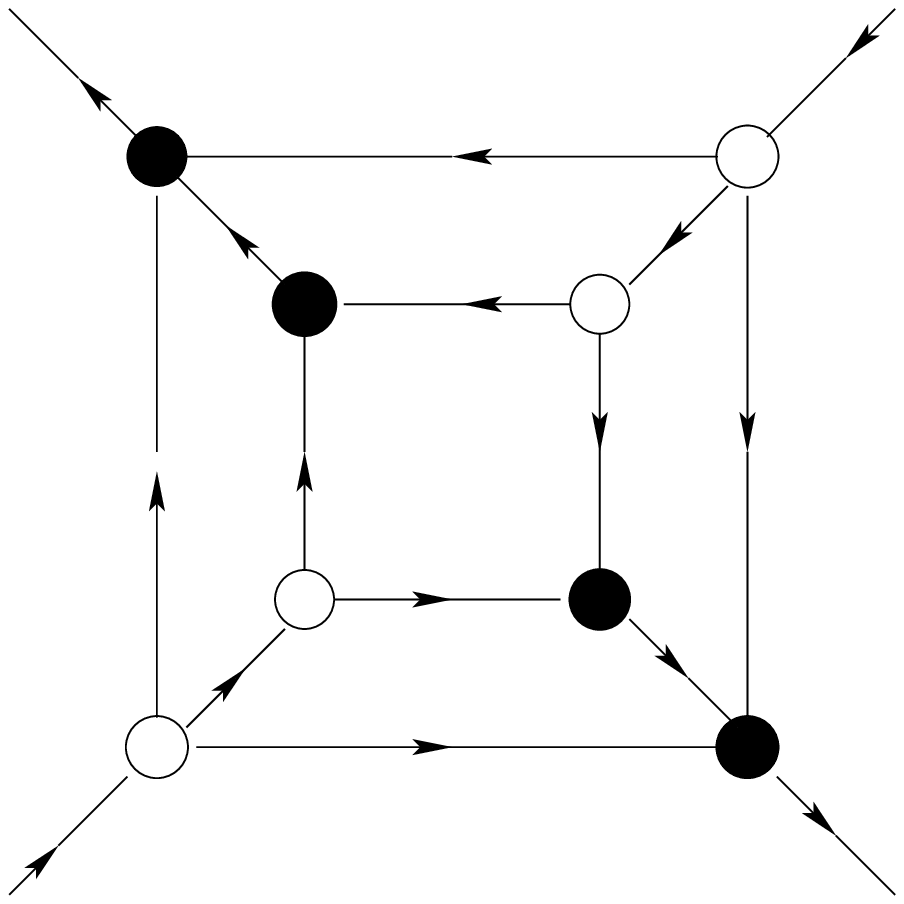}}}}
   \hspace{-.2cm}=\:
  &M_4^{\mbox{\tiny tree}}
 \mathcal{F}(\varepsilon,\,\bar{\varepsilon})\,
 \Omega_{4}(\zeta),\qquad
 \Omega_{4}(\zeta)\:\equiv\:
  \bigwedge_{i=1}^4\,d\zeta_{i,i+1}\zeta_{i,i+1}^{2\varepsilon_{i,i+1}-1}
  (1-\zeta_{i,i+1})^{-2(\varepsilon_{i,i+1}+\bar{\varepsilon}_{i,i+1})},
 \end{split}
\end{equation}\end{widetext}
where $\mathcal{F}(\varepsilon,\,\bar{\varepsilon})$ is the 
parameter-dependent factor in (\ref{eq:M4tree}), up to the substitution
$\varepsilon_{i,i+1}\,\longrightarrow\,
 \varepsilon_{i,i+1}+\bar{\varepsilon}_{i,i+1}$, with 
$\bar{\varepsilon}_{i,i+1}$ and $\varepsilon_{i,i+1}$ being the deformation
parameters related to the internal legs of the inner and outer box 
respectively on the l.h.s., the variables 
$\zeta_{i,i+1}$'s parametrizing the on-shell diagram phase-space are
related to the $z_{i,i+1}$'s left over after the gluing procedures by
a simple change of variable. Notice that equation (\ref{eq:M41L}) 
provide a representation for the one-loop integrand of the four-particle 
amplitude with fixed external gluon states in $\mathcal{N}\,=\,4$ SYM as 
well as it can be seen as the contribution to the one-loop integrand 
carrying information about one of the leading singularities in pure 
Yang-Mills.

Some comments are now in order. Firstly, setting all the parameters to zero,
one obtains the $\bigwedge_{i=1}^4\,d(\log{\zeta_{i,i+1}})$ form for the 
above integrand, as it should be for UV-finite contributions. Secondly,
the $\Omega_4(\zeta)$ in (\ref{eq:M41L}) can be seen as a four-form
with branch points at $\zeta_{i,i+1}\,=\,0,\infty,\,1$. Thus, the 
introduction of our helicity deformations opens up the poles naturally
present in the integrand producing branch cuts as well as new
singularities ($\zeta_{i,i+1}\,=\,1$) not present in the undeformed 
integrand. The $4$-form $\Omega_4$ can be thought of as a holomorphic 
multi-valued $4$-form on the complement of a branch locus in $\mathbb{C}^4$.
In order to extract physical information from (\ref{eq:M41L}), one has
to identify the correct integration path and how to correctly treat the
branch locus.
We propose that the contour of integration which allows to compute 
(a UV-finite contribution to) the four-particle amplitude in Lorentz
signature is given by
\begin{equation}\label{eq:IntPath}
 \Gamma\:=\:
  \left\{
   \zeta\in\mathbb{C}^4\,\Big|\,
   -\frac{s}{u}\zeta_{12}\zeta_{34}-
    \frac{t}{u}\zeta_{23}\zeta_{41}\:=\:
    \frac{\zeta_{i,i+1}}{\zeta_{i,i+1}^{\star}}, \forall\,i
   \right\},
\end{equation}
$\zeta_{i,i+1}^{\star}$ being the complex conjugate of $\zeta_{i,i+1}$.
Notice that $\Gamma$ contains the right symmetries and it can be checked 
bottom-up by relating the $\zeta$-parametrization of the phase-space to the 
off-shell loop momentum.
The path (\ref{eq:IntPath}) implies that all the $\zeta$'s have the 
same phase (up to $k\,\pi$),
so one can adopt a convenient parametrization
of the path in terms of a phase and three real variables. Since our path
crosses branch points, as it is usually done in complex analysis, we split
the integral into several ones, keeping track of the $e^{i\pi}$ factors 
generated by the hidden logarithms in \eqref{eq:M41L} ($z^w=e^{w\log(z)}$) 
when moving from one region to the other. After some manipulations, 
the result can be obtained by computing integrals of the form 
\begin{widetext}
\begin{equation}\label{eq:Integral}
 I_4\:=\:\oint_{|\omega|=1}\frac{d\omega}{\omega}
  \omega^{2\sum_{i}\varepsilon_{i,i+1}}
  \int\bigwedge_{i\,=\,1}^4\,dx_{i,i+1}x_{i,i+1}^{2\varepsilon_{i,i+1}-1}
  \left(1-\omega x_{i,i+1}\right)^{-2(\varepsilon_{i,i+1}
   +\bar{\varepsilon}_{i,i+1})}
  \delta\left(1+\frac{s}{u}x_{12}x_{34}+\frac{t}{u}x_{23}x_{41}\right),
\end{equation}
\end{widetext}
where the range for the $x_{i,i+1}$ is from $0$ to $\infty$ or from
$-\infty$ to $0$ depending on the region. After using the $\delta$-function
to solve for, again depending on the region, either $x_{34}$ or $x_{41}$, 
one is left with an integral whose divergent part can be extracted using two
Mellin-Barnes transforms on the terms involving the variable we solved for. 
In spirit it is the same calculation one can use when dimensionally
regularizing the box integral, but the details are a bit more involved and we
postpone a detailed discussion to \cite{Benincasa:2014wp1}.

The result for the integration of $\Omega_{4}$ (which we will refer to as 
$I_4$) when the $\epsilon_{i,i+1},\bar\epsilon_{i,i+1}$ are taken to be 
proportional to $\epsilon$ is an expression that resembles very much the 
results obtained with dimensional or analytic regularization. In order to 
display more explicitly this behavior, we take particular values of these 
eight parameters. For instance, when 
$\bar\epsilon_{i,i+1}=\epsilon_{i,i+1}$ and 
$\epsilon_{12}=\epsilon_{23}=\epsilon$,
$\epsilon_{34}=\epsilon_{41}=2\epsilon$, up to some convention-dependent
normalization factor $I_4$ is equal to:
\begin{widetext}
\begin{equation}\label{eq:Ieg}
 \frac{1}{\epsilon^2}-3\frac{\log\left(-\frac{s}{u}\right)+\log\left(-\frac{t}{u}\right)}{2\epsilon}
 +\frac{1}{2}\log\left(-\frac{s}{u}\right)^2+\frac12\log\left(-\frac{t}{u}\right)^2+
 2\log\left(-\frac{s}{u}\right)\log\left(-\frac{t}{u}\right)-\frac{56\pi^2}{3}\,,
\end{equation}
\end{widetext}
where we are working in the kinematical region $s<0$, $t<0$. The result 
above is symmetric under the label exchange $s\rightleftharpoons t$, but 
this is just an artifact of our choice of regularization parameters: $I_4$ 
posseses the same symmetries as the l.h.s. of (\ref{eq:M41L}), 
and as such is 
invariant under relabelings $2\rightleftharpoons4$ and 
$1\rightleftharpoons3$. For generic values of the regularization parameters,
$I_4$ will not be invariant under $s\rightleftharpoons t$, as it happens for
dimensional regularization. This is consistent with the fact that 
introducing different helicity deformations for the scattering particles 
breaks some of these discrete symmetries. Another difference with 
dimensional regularization that we can observe in \eqref{eq:Ieg} is that
the third Mandelstam variable $u$ is playing the role of 
``renormalization scale''. Of course our regularization procedure remains 
four-dimensional all along and we do not introduce any extra dimensionful 
scale in the problem.

\section{Conclusion and outlook}\label{sec:Concl}

In this paper we propose a general (theory-independent) scheme for 
regularizing on-shell forms, which allows to have well-defined integrals
on the Lorentz sheet. Our idea comes from two basic considerations. 
First, the fundamental objects for the on-shell 
representation of a scattering amplitude are the three-particle ones, 
whose coupling constant dimensionality $[\kappa]$ 
allows to classify theories.
The dimension of the three-particle coupling constant depends on the
helicities, so that fixing it restricts the possible states which
can interact in a given three-particle amplitude. Secondly,
the on-shell construction breaks locality at intermediate steps, {\it i.e.}
a single on-shell diagram can show poles which are not in the amplitude
it contributes to. With these considerations in mind, we propose to perform
a deformation of the helicity space on the three-particle amplitudes, in
such a way that its dimensionality is not changed. This constraint
allows us to keep a four-dimensional framework as well as to remain in a
given class of theories at fixed $[\kappa]$ but extending it to 
{\it unphysical} values of the helicities. As a consequence of this 
extension, locality gets broken. A regularization scheme typically breaks
some features of a theory, which is the price one has to pay to be able
to have well-defined quantities. Given that, at least at intermediate
steps, the on-shell construction breaks locality, it can be more suitable
in such a framework to introduce a regularization scheme which performs such
a breaking (in a controlled way), rather than giving up any other feature or
symmetry. We discuss the locality breaking in our scheme (even if a more
detalied discussion will appear in \cite{Benincasa:2014wp1}),
arguing that it allows to associate the relevant physical quantities to 
poles in the parameter space. 
More precisely, even at tree level, one can read off
the collinear behavior of an amplitude as well as the underformed on-shell
diagrams from poles in the deformation parameter space. At loop level, the 
IR singularities are reflected also as poles in the parameter space, in
a similar fashion to what happens in dimensional regularization. Our 
computation of a UV-finite contribution to a one-loop four-particle 
amplitude reveals both the correct IR structure and the correct functional
structure at order $\mathcal{O}(1)$. It is interesting that at
integrand level, our deformation maps the four-form
$\bigwedge_{i=1}^4d(\log{\zeta_{i,i+1}})$ into another four-form which
is nothing but the integrand of the Euler beta-function, even if
then the integration contour $\Gamma$ complicates the branch cut structure.
As already said in the text, our computation has been performed in a sort
of {\it brute force} way, but it would be interesting to exploit the power
of the hypergeometric function theory to have a cleaner and somehow more
natural way to treat the integration and the branch locus.

In order to perform the integration directly in the {\it on-shell variables}
we needed to find a contour of integration implementing the Lorentz sheet.
We proceded by inspection but still we do not have a general, first principle
critirium to define such contours. Finding such a criterium, with the hope
of extending it to higher loops, still remains an open question, even if
we consider instructive to have shown both the shape of the Lorentz sheet
from the on-shell perspective and have performed directly the integration.

\acknowledgments

\section{Acknowledgments}

We would like to thank Riccardo Argurio, Andreas Brandhuber, 
Simon Caron-Huot, Pierpaolo Mastrolia,  Micha Moskovic, Diego Redigolo, 
Agust{\'i}n Sabio Vera, Alberto Salvio, Juan Jos{\'e} Sanz-Cillero, 
Gabriele Travaglini, Jaroslav Trnka for valuable discussions. 
P.B. and D.G. are supported in part by Plan 
Nacional de Altas Energ{\'i}as (FPA2011-25948 and FPA2012-32828), Spanish MICINN Consolider-Ingenio 2010 Program CPAN (CSD2007-00042), Comunidad de 
Madrid HEP-HACOS S2009/ESP-1473 and the Spanish MINECO's Centro de
Excelencia Severo Ochoa Programme under grants SEV-2012-0234 and 
SEV-2012-0249. The work of E.C. is supported in part by the Belgian Fonds de la Recherch Fondamentale Collective (grant 2.4655.07), the Belgian Institut
Interuniversitaire des Sciences Nucl\'eaires (grants 4.4511.06 and 
4.4514.08), 
by the ``Communaut\'e Francaise de Belgique'' through the ARC program
and by the ERC through the ``SyDuGraM'' Advanced Grant. E.C. would also like to acknowledge the MP1210 COST Action and hospitality of the IFT.
D.G. is a {\it La Caixa-Severo Ochoa Scholar} and thanks La Caixa foundation
for financial support.

\bibliography{amplitudesrefs}

\begin{thebibliography}{14}
\expandafter\ifx\csname natexlab\endcsname\relax\def\natexlab#1{#1}\fi
\expandafter\ifx\csname bibnamefont\endcsname\relax
  \def\bibnamefont#1{#1}\fi
\expandafter\ifx\csname bibfnamefont\endcsname\relax
  \def\bibfnamefont#1{#1}\fi
\expandafter\ifx\csname citenamefont\endcsname\relax
  \def\citenamefont#1{#1}\fi
\expandafter\ifx\csname url\endcsname\relax
  \def\url#1{\texttt{#1}}\fi
\expandafter\ifx\csname urlprefix\endcsname\relax\def\urlprefix{URL }\fi
\providecommand{\bibinfo}[2]{#2}
\providecommand{\eprint}[2][]{\url{#2}}

\bibitem[{\citenamefont{Arkani-Hamed et~al.}(2012)\citenamefont{Arkani-Hamed,
  Bourjaily, Cachazo, Goncharov, Postnikov et~al.}}]{ArkaniHamed:2012nw}
\bibinfo{author}{\bibfnamefont{N.}~\bibnamefont{Arkani-Hamed}},
  \bibinfo{author}{\bibfnamefont{J.~L.} \bibnamefont{Bourjaily}},
  \bibinfo{author}{\bibfnamefont{F.}~\bibnamefont{Cachazo}},
  \bibinfo{author}{\bibfnamefont{A.~B.} \bibnamefont{Goncharov}},
  \bibinfo{author}{\bibfnamefont{A.}~\bibnamefont{Postnikov}},
  \bibnamefont{et~al.} (\bibinfo{year}{2012}), \eprint{1212.5605}.

\bibitem[{\citenamefont{Huang and Wen}(2014)}]{Huang:2013owa}
\bibinfo{author}{\bibfnamefont{Y.-T.} \bibnamefont{Huang}} \bibnamefont{and}
  \bibinfo{author}{\bibfnamefont{C.}~\bibnamefont{Wen}},
  \bibinfo{journal}{JHEP} \textbf{\bibinfo{volume}{1402}}, \bibinfo{pages}{104}
  (\bibinfo{year}{2014}), \eprint{1309.3252}.

\bibitem[{\citenamefont{Huang et~al.}(2014)\citenamefont{Huang, Wen, and
  Xie}}]{Huang:2014xza}
\bibinfo{author}{\bibfnamefont{Y.-t.} \bibnamefont{Huang}},
  \bibinfo{author}{\bibfnamefont{C.}~\bibnamefont{Wen}}, \bibnamefont{and}
  \bibinfo{author}{\bibfnamefont{D.}~\bibnamefont{Xie}} (\bibinfo{year}{2014}),
  \eprint{1402.1479}.

\bibitem[{\citenamefont{Elvang et~al.}(2014)\citenamefont{Elvang, Huang,
  Keeler, Lam, Olson et~al.}}]{Elvang:2014fja}
\bibinfo{author}{\bibfnamefont{H.}~\bibnamefont{Elvang}},
  \bibinfo{author}{\bibfnamefont{Y.-t.} \bibnamefont{Huang}},
  \bibinfo{author}{\bibfnamefont{C.}~\bibnamefont{Keeler}},
  \bibinfo{author}{\bibfnamefont{T.}~\bibnamefont{Lam}},
  \bibinfo{author}{\bibfnamefont{T.~M.} \bibnamefont{Olson}},
  \bibnamefont{et~al.} (\bibinfo{year}{2014}), \eprint{1410.0621}.

\bibitem[{\citenamefont{Arkani-Hamed and Trnka}(2013)}]{Arkani-Hamed:2013jha}
\bibinfo{author}{\bibfnamefont{N.}~\bibnamefont{Arkani-Hamed}}
  \bibnamefont{and} \bibinfo{author}{\bibfnamefont{J.}~\bibnamefont{Trnka}}
  (\bibinfo{year}{2013}), \eprint{1312.2007}.

\bibitem[{\citenamefont{Bourjaily et~al.}(2013)\citenamefont{Bourjaily,
  Caron-Huot, and Trnka}}]{Bourjaily:2013mma}
\bibinfo{author}{\bibfnamefont{J.~L.} \bibnamefont{Bourjaily}},
  \bibinfo{author}{\bibfnamefont{S.}~\bibnamefont{Caron-Huot}},
  \bibnamefont{and} \bibinfo{author}{\bibfnamefont{J.}~\bibnamefont{Trnka}}
  (\bibinfo{year}{2013}), \eprint{1303.4734}.

\bibitem[{\citenamefont{Ferro et~al.}(2013)\citenamefont{Ferro, Lukowski,
  Meneghelli, Plefka, and Staudacher}}]{Ferro:2012xw}
\bibinfo{author}{\bibfnamefont{L.}~\bibnamefont{Ferro}},
  \bibinfo{author}{\bibfnamefont{T.}~\bibnamefont{Lukowski}},
  \bibinfo{author}{\bibfnamefont{C.}~\bibnamefont{Meneghelli}},
  \bibinfo{author}{\bibfnamefont{J.}~\bibnamefont{Plefka}}, \bibnamefont{and}
  \bibinfo{author}{\bibfnamefont{M.}~\bibnamefont{Staudacher}},
  \bibinfo{journal}{Phys. Rev. Lett.} \textbf{\bibinfo{volume}{110}},
  \bibinfo{pages}{121602} (\bibinfo{year}{2013}), \eprint{1212.0850}.

\bibitem[{\citenamefont{Ferro et~al.}(2014{\natexlab{a}})\citenamefont{Ferro,
  Lukowski, Meneghelli, Plefka, and Staudacher}}]{Ferro:2013dga}
\bibinfo{author}{\bibfnamefont{L.}~\bibnamefont{Ferro}},
  \bibinfo{author}{\bibfnamefont{T.}~\bibnamefont{Lukowski}},
  \bibinfo{author}{\bibfnamefont{C.}~\bibnamefont{Meneghelli}},
  \bibinfo{author}{\bibfnamefont{J.}~\bibnamefont{Plefka}}, \bibnamefont{and}
  \bibinfo{author}{\bibfnamefont{M.}~\bibnamefont{Staudacher}},
  \bibinfo{journal}{JHEP} \textbf{\bibinfo{volume}{1401}}, \bibinfo{pages}{094}
  (\bibinfo{year}{2014}{\natexlab{a}}), \eprint{1308.3494}.

\bibitem[{\citenamefont{Beisert et~al.}(2014)\citenamefont{Beisert, Broedel,
  and Rosso}}]{Beisert:2014qba}
\bibinfo{author}{\bibfnamefont{N.}~\bibnamefont{Beisert}},
  \bibinfo{author}{\bibfnamefont{J.}~\bibnamefont{Broedel}}, \bibnamefont{and}
  \bibinfo{author}{\bibfnamefont{M.}~\bibnamefont{Rosso}}, \bibinfo{journal}{J.
  Phys. A} \textbf{\bibinfo{volume}{47}}, \bibinfo{pages}{365402}
  (\bibinfo{year}{2014}), \eprint{1401.7274}.

\bibitem[{\citenamefont{Ferro et~al.}(2014{\natexlab{b}})\citenamefont{Ferro,
  Lukowski, and Staudacher}}]{Ferro:2014gca}
\bibinfo{author}{\bibfnamefont{L.}~\bibnamefont{Ferro}},
  \bibinfo{author}{\bibfnamefont{T.}~\bibnamefont{Lukowski}}, \bibnamefont{and}
  \bibinfo{author}{\bibfnamefont{M.}~\bibnamefont{Staudacher}},
  \bibinfo{journal}{Nucl. Phys. B} \textbf{\bibinfo{volume}{889}},
  \bibinfo{pages}{192} (\bibinfo{year}{2014}{\natexlab{b}}),
  \eprint{1407.6736}.

\bibitem[{\citenamefont{Benincasa and Cachazo}(2007)}]{Benincasa:2007xk}
\bibinfo{author}{\bibfnamefont{P.}~\bibnamefont{Benincasa}} \bibnamefont{and}
  \bibinfo{author}{\bibfnamefont{F.}~\bibnamefont{Cachazo}}
  (\bibinfo{year}{2007}), \eprint{0705.4305}.

\bibitem[{\citenamefont{Benincasa and Conde}(2012)}]{Benincasa:2011pg}
\bibinfo{author}{\bibfnamefont{P.}~\bibnamefont{Benincasa}} \bibnamefont{and}
  \bibinfo{author}{\bibfnamefont{E.}~\bibnamefont{Conde}},
  \bibinfo{journal}{Phys.Rev.} \textbf{\bibinfo{volume}{D86}},
  \bibinfo{pages}{025007} (\bibinfo{year}{2012}), \eprint{1108.3078}.

\bibitem[{\citenamefont{Benincasa et~al.}(2014)\citenamefont{Benincasa, Conde,
  and Gordo}}]{Benincasa:2014wp1}
\bibinfo{author}{\bibfnamefont{P.}~\bibnamefont{Benincasa}},
  \bibinfo{author}{\bibfnamefont{E.}~\bibnamefont{Conde}}, \bibnamefont{and}
  \bibinfo{author}{\bibfnamefont{D.}~\bibnamefont{Gordo}},
  \bibinfo{journal}{{\it in preparation}}  (\bibinfo{year}{2014}).

\bibitem[{\citenamefont{Britto et~al.}(2005)\citenamefont{Britto, Cachazo,
  Feng, and Witten}}]{Britto:2005aa}
\bibinfo{author}{\bibfnamefont{R.}~\bibnamefont{Britto}},
  \bibinfo{author}{\bibfnamefont{F.}~\bibnamefont{Cachazo}},
  \bibinfo{author}{\bibfnamefont{B.}~\bibnamefont{Feng}}, \bibnamefont{and}
  \bibinfo{author}{\bibfnamefont{E.}~\bibnamefont{Witten}},
  \bibinfo{journal}{Phys. Rev. Lett.} \textbf{\bibinfo{volume}{94}}
  (\bibinfo{year}{2005}), \eprint{0501052}.

\end{thebibliography}

\end{document}